# Coarse Grain Molecular Dynamics Simulation of Fibrin Polymerization

Sumith Yesudasan and Rodney D. Averett*

*School of Chemical and Biological Engineering, University of Georgia, 597 D.W. Brooks Drive, Athens, GA 30602*

*Corresponding author email: raverett@uga.edu
Telephone: 706-542-0863

**Abstract**

Studies suggests that patients with deep vein thrombosis and diabetes often have hyper coagulable blood plasma leading to higher chances of forming thromboembolisms by the rupture of blood clots, which may lead to stroke and death. Despite the advances in the field of blood clot formation and lysis research, the change in mechanical properties and its implication into the formation of thromboembolisms in platelet poor plasma is poorly understood. In this paper, we present a new computational method to simulate fibrin clot formation using molecular simulations. With an effective combination of reactive molecular dynamics concept and coarse graining principle, we have utilized the reactive coarse grain molecular dynamics to predict the complex network formation of fibrin clots and the branching of the fibrins. The heavy 340 kDa fibrinogen is converted into a simple spring-bead coarse grain system with 9 beads, and using our customized reactive potentials, we simulated the formation of the fibrin clot. Thus, formed fibrin clot agrees with the experimental results qualitatively, and to our best knowledge this is the first kind of molecular polymerization study of fibrin clot which can lead to improve our understanding about blood clot formation and its relationship with mechanical properties.

## Introduction

Deep vein thrombosis is the formation of a blood clot (thrombus) within a deep vein most commonly in the legs, which can lead to wide range of complications[1]. In the event of rupturing of the thrombus, it may travel to the lungs and leads to pulmonary embolism which can be fatal[2, 3] if left untreated. Vascular endothelial damage, stasis of blood flow, and hypercoagulability of blood is found to be directly related to the risk factors of deep vein thrombosis[4-7] and pulmonary embolism[8-11]. The venous thromboembolism is considered as an epidemic[12-16] and various thrombolytic therapies are suggested by the clinicians to control it after a typical surgery[17-27]. The characterization of the mechanical properties of the thrombo-emboli plays an important role in developing thrombolytic therapies[28], and towards this goal a number of studies have been done recently[29-33]. Despite of all these developments in this area, the computational modeling is not well developed. An accurate computational model based on the molecular information will be helpful in predicting the mechanical behavior of the blood clots under various pathogenic states and also can be helpful to design better thrombolytic therapies. In this work, towards this goal we will lay the foundations of simulating fibrin clot using molecular dynamics simulations.

The fibrinogen is a big soluble glycoprotein which plays a major role in the event of blood clot, by forming network of fibrins which is the basis of a blood clot. It took almost 50 years to get the accurate structural representation of the fibrinogen molecule[34]. Previously, the tri-nodular structure of the fibrinogen molecule was determined by various methods including electron micrographs[35]. Using the molecular models of fibrinogen and thrombin, an enzyme which cleaves the fibrinopeptides of fibrinogen and thereby initiating the polymerization, studies were conducted to understand their interactions[36]. Towards the simulation of blood clot and blood flow, there exists a few studies with discrete particles[37] and multiscale models[38-41]. There are studies aimed at understanding the adsorption behavior of the fibrinogen with different types of surfaces[42] which can be helpful in understanding immunological response[43].

Even after this many computational studies related to blood clot, there exists no models which can accurately predict the formation of the blood clot, its mechanical behavior, adsorption, and lysis events based on the molecular information. They are mainly based on the empirical rate of clot formation information modeled as first order differential equation. A multiscale model which can leverage the molecular information to the macroscale clot mechanical properties can elucidate the thrombus rupture mechanism and thereby we can aim at developing new targeted thrombolytic therapies. As the first step towards this, we have developed a highly coarse grain model of fibrinogen and then this solvent free model along with the modified reactive molecular dynamics potentials we simulate the clot formation at molecular level. The force field parameters are optimized using our custom written computer code, and we have observed some important events in the fibrin clot such as the continuous long strand formation, and fibrin branching and cross-linking.

## Coarse Grain Model Development

A coarse grain model with minimal number of beads is preferred for fibrinogen to avoid the computational overhead, at the same time having enough beads and springs to maintain the flexibility and extensibility of the molecule. With this in mind, we used a shape based coarse graining approach and have divided the human fibrinogen (RCSB 3GHG[34]) molecule into nine fragments and estimated their molecular mass and converted them into beads separated by harmonic springs of 5.75 nm in length. Figure 1a shows the molecular model of human fibrinogen[34] with α, β, and γ-chains along with the D-region, E-region and the coiled region (C-region). The mass partitioning procedure is shown in Fig. 1b, and the final coarse grain model is shown in Fig. 1c with C-type, D-type and E-type beads. The type-C beads weigh 20 kDa, type-D weighs 32.5 kDa and type-E weighs 25 kDa respectively. Additionally, type-T beads (not shown in figure) are considered with 20 kDa mass which functions similar to thrombin. The actual molecular weights of thrombin (RCSB 1PPB[44]) is 34.3 kDa and factor XIII (RCSB 1GGU[45]) is 166 kDa; however to enhance the mobility of T-bead in a solvent-free environment, we arbitrarily selected it as 20 kDa.

To maintain the connectivity and simulate the elastic behavior, we used a harmonic bond potential (Eqn. 1) which connects bead-D to bead-C, bead-C to bead-C and bead-C to bead-E.

$$V_{bond} = k_b(r - r_0)^2 \quad (1)$$

Here, $2k_b$ is the stiffness of the spring, $r_0$ is the equilibrium distance of the bonds, and $r$ is the spacing between the beads in the bond. The value of the linear spring stiffness is estimated from all atom molecular dynamics simulation[46] of a single fibrinogen molecule as $0.75\ N/m$. The equilibrium bond distance is considered as $5.75\ nm$ between all types of beads, which accounts for the actual fibrinogen length of 46 nm. For maintaining the shape and to avoid worm-like chain behavior of the fibrinogen molecule we applied an angle harmonic potential (Eq. 2) between every adjacent bond.

$$V_{angle} = k_a(\theta - \theta_0)^2 \quad (2)$$

Here, $k_a$ is the potential stiffness, which controls the flexibility of the bonds and $\theta_0$ is the equilibrium angle in this case taken as 180 deg.

Arbitrarily, we assigned $10.432\ \frac{kJ}{mol\ rad^2}$ as the spring stiffness, which makes the molecule neither too stiff nor too flexible. This value can be and has to be optimized to reflect the bending stiffness of the molecule; nevertheless, this study was aimed at the polymerization and hence we chose it arbitrarily.

All the beads interact with each other using a standard shifted truncated LJ potential (Eqn. 3) with a cutoff radius of 50 nm. The potential and its negative derivative (force) goes to zero smoothly when $r$ approaches cutoff radius ($r_c$). The depth of the potential well ($\varepsilon$) is $0.23838\ kJ/mol$ (same as argon) and $\sigma = 4\ nm$ is used between all beads. However, $\sigma$ is changed to $2.5\ nm$ between C-C bead interaction, and to $1\ nm$ between beads T and E.

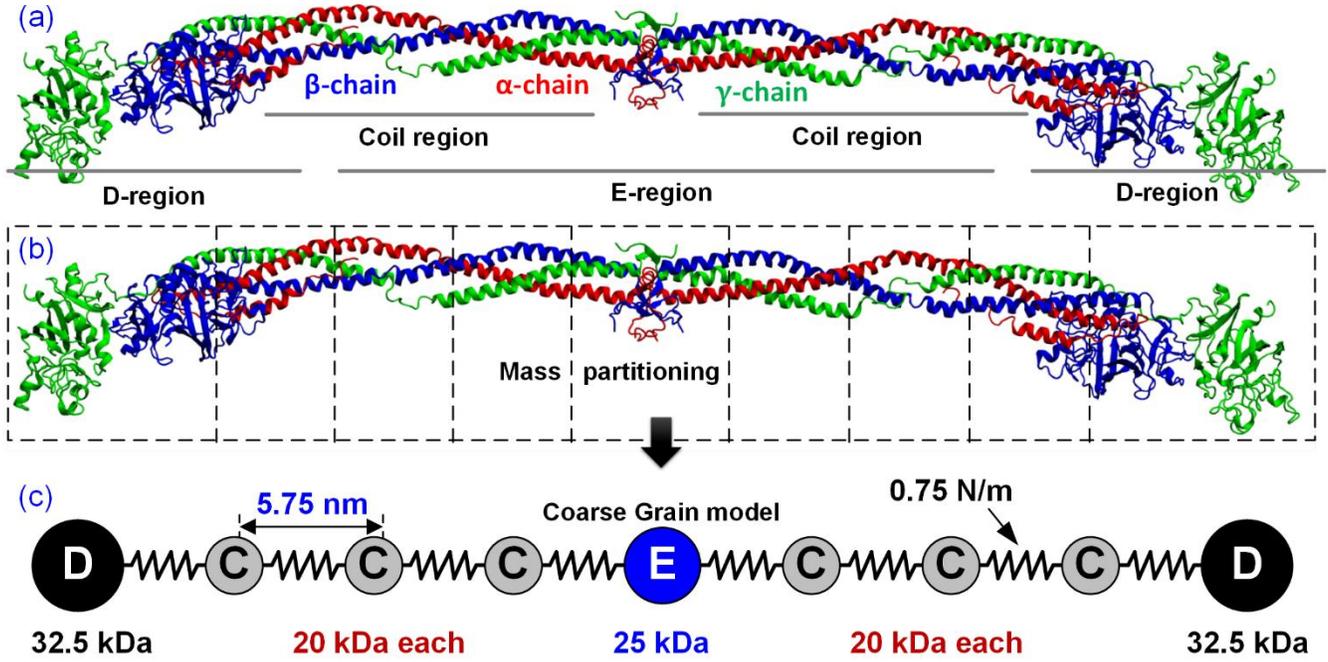

**Figure 1:** Coarse graining steps of fibrinogen molecule. (a) Molecular model of human fibrinogen with α-chain, β-chain and γ-chain in cartoon representation using VMD[47]. (b) Partitioning of the fibrinogen molecule based on the mass. (c) Coarse grain model of fibrinogen with mass, bond length and type of different beads.

This is to avoid unnecessary calculation of LJ potential and harmonic potential at the same time, which may lead to a bouncing off (repulsive LJ) and attraction (harmonic). Here bead-T represents the thrombin bead, and its role is discussed in the next section.

$$V_{LJ} = 4\varepsilon\left[\left(\frac{\sigma}{r}\right)^{12} - \left(\frac{\sigma}{r}\right)^{6}\right] - 4\varepsilon(r - r_c)\left[\frac{6\sigma^6}{r_c^7} - \frac{12\sigma^{12}}{r_c^{13}}\right] - 4\varepsilon\left[\frac{\sigma^{12}}{r_c^{12}} - \frac{\sigma^6}{r_c^6}\right] \quad (3)$$

The Eqn. 3 represents the shifted LJ potential, where $r$ is the inter atomic/bead spacing.

## Polymerization force field development

In this section, we will explain the general pathway of fibrin clot formation, the molecular mechanism behind it, development of the polymerizing force field, and the development of the computer code used for simulating the fibrin clot.

### Mechanism of Fibrin polymerization

The fibrinogen is converted to fibrin monomer upon the cleaving of fibrinopeptides in the E-region by activated thrombin (factor IIa), and remains bound to the fibrinogen molecule[48-50]. These fibrin monomers will start forming protofibrils of two fibrins, which starts polymerizing as long strands. These long strands in the presence of factor XIIIa, will aggregate laterally and will form a stable fibrin clot. The factor XIIIa also binds into fibrin clot[51, 52] and completes the coagulation cascade[53]. The various steps involved in this process is explained graphically in Fig. 2 and a detailed description of the chain of events and various stages of coagulation cascade are in the literature[53, 54]. Figure 2a shows the schematic of fibrinogen molecule along with the thrombin beads suspended in the plasma system. An exact molecular simulation of this bio-chemical reaction at atomic scale is very challenging to simulate, and hence we rely on specially developed solvent free coarse grain force field to demonstrate the clotting process. Figure 2b shows the phase-I of the polymerization, in which the thrombin cleaves the fibrinopeptides of the fibrinogen and makes fibrin monomers. In reactive coarse grain molecular simulations (RCGMD), we simulate this by bonding type-T beads to type-E beads. In the next phase, the fibrin monomers will start polymerizing with each other, type-D beads bonding to type-T beads with specific rules, and this similar to the thrombin binding on D-region[55]. This process is referred to as phase-II polymerization. The time evolution of this simulation will demonstrate the long fibrin strand formation and lateral branching and aggregation of the fibrin polymers as shown in Fig. 2c.

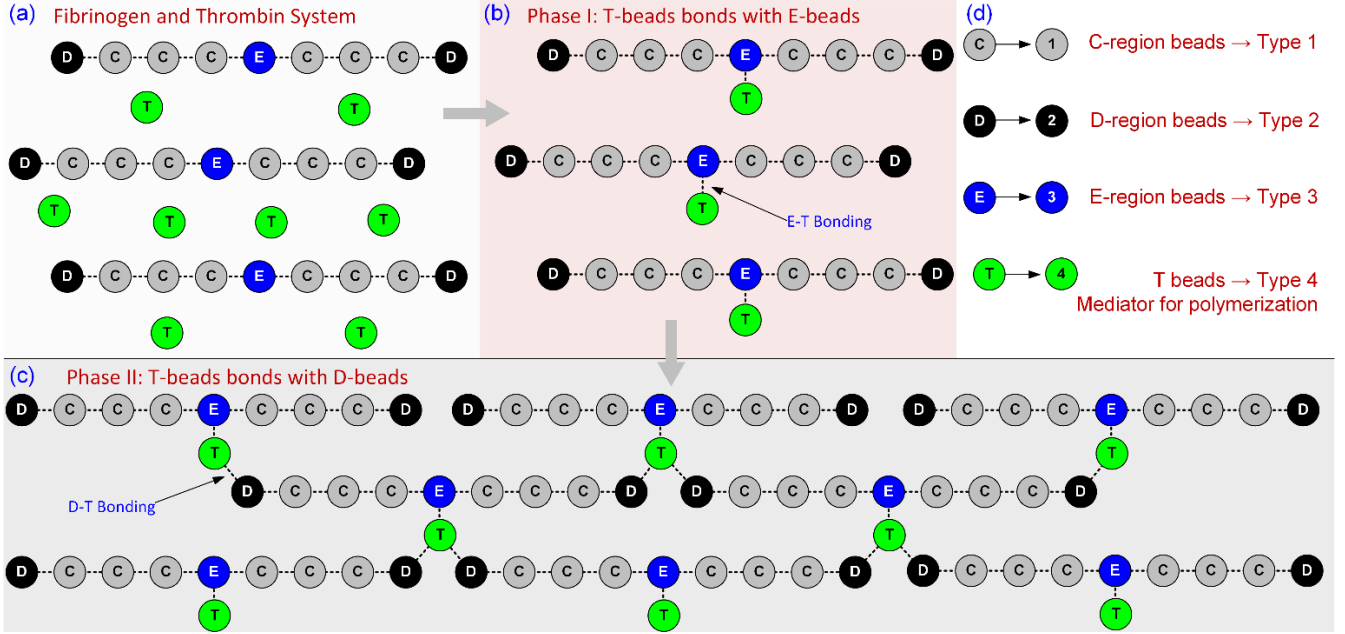

**Figure 2:** A conceptual representation of the fibrin polymerization. (a) Fibrinogen molecules and thrombin like beads. (b) The T-beads attaches to E-beads, which is analogous to the process of thrombin (factor IIa) cleaving the fibrinopeptides (fpA and fpB). (c) In this stage, D-beads bonds with T-beads, which is analogous to the D-regions bonding to E-region in clotting cascade. (d) Molecule code naming conventions for various beads. Note that T-beads (Type 4) is just a mediator to form bonds in simulations, which has similar functions of thrombin.

The RCGMD basis of the fibrin clotting process can be classified into phase-I and phase-II, in which the common types of potentials are used along with rule based switching functions to enable the clotting simulation.

**Phase I: Type 3-4 bonding**

RCGMD force field consists of three potentials, an attractive, a repulsive and a bonding potential as shown in equations 4, 5 and 6 respectively. Here, $k_1$ is the common constant for attractive and repulsive potential, $k_2$ is the spring constant for bonding potential, $r_{bond}$ is the bond distance between bead types 3 and 4, and $r$ is the inter-bead distance. $r_{bond}$ is kept as $3.5\ nm$ for all our simulations.

$$V_{attract} = -k_1/(r - r_{bond})^2 \quad (4)$$

$$V_{repel} = k_1/r \quad (5)$$

$$V_{Bonding} = k_2(r - r_{bond})^2 \quad (6)$$

These potentials are combined together using heavyside functions as shown in eqn. 7 to get the phase-I RCGMD potential. Heavyside function (eqn. 13) acts as a switch, which returns a value of 1 for positive numbers, 0 for negative and 0.5 for zero. Unit of $k_1$ in Eqn. 4 is $kJ.nm^2/mol$ and $kJ.nm/mol$ in Eqn. 5, and $k_2$ is $kJ/(mol.nm^2)$. Despite of the unit differences, we chose a common constant for repulsive and attractive potential as $k_1$ so as to reduce the optimization effort.

$$V_{rcgmd-I} = (\Theta_1\Theta_4)V_{attract} + (\Theta_2\Theta_5)V_{Bonding} + (\Theta_3 + \Theta_2\Theta_4)V_{repel} \quad (7)$$

$$\Theta_1 = \Theta(0.9 - N_{34}) \quad (8)$$

$$\Theta_2 = \Theta([N_{34} - 0.9]/[1.1 - N_{34}]) \quad (9)$$

$$\Theta_3 = \Theta(N_{34} - 1.9) \quad (10)$$

$$\Theta_4 = \Theta(r^2 - r^2_{threshold}) \quad (11)$$

$$\Theta_5 = \Theta(r_{threshold}^2 - r^2) \qquad (12)$$

$$\Theta(t) = \begin{cases} 0, & \text{for } t < 0 \\ 1/2, & \text{for } t = 0 \\ 1, & \text{for } t > 0 \end{cases} \qquad (13)$$

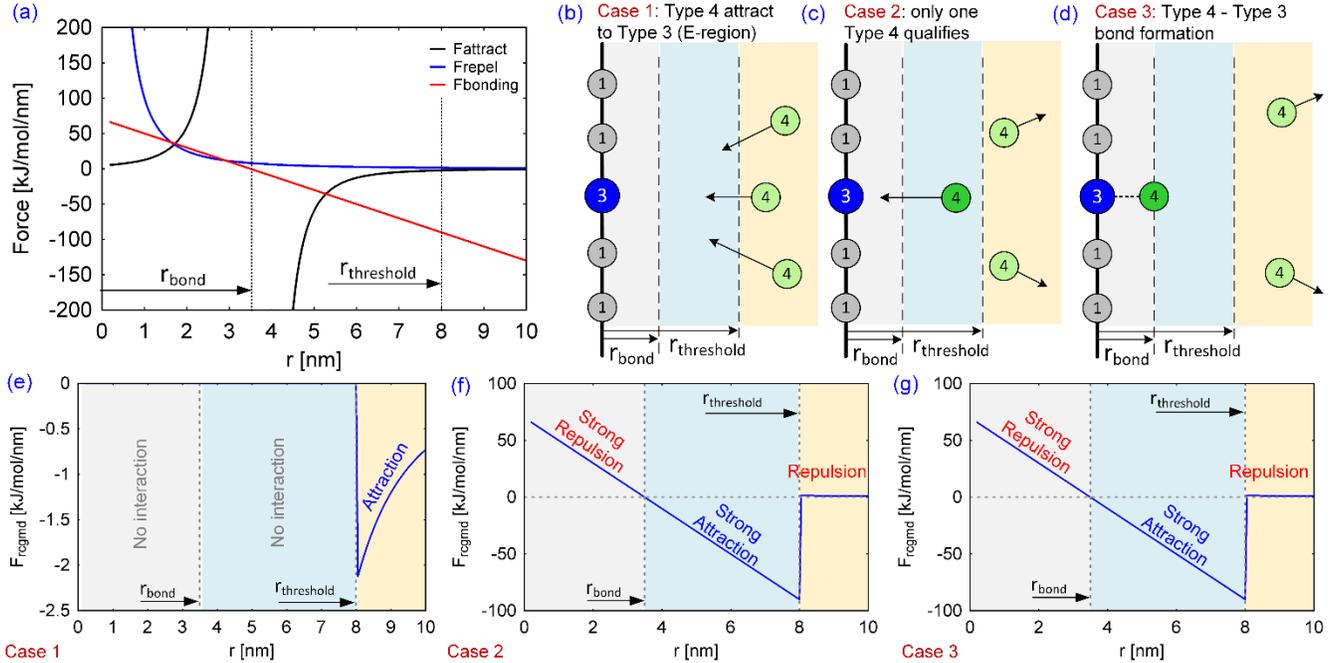

**Figure 3:** Different types of potentials that we used to develop the reactive force field. (a) It consists of a shifted attractive potential, a repulsive potential and a harmonic potential. (b) when no type-4 beads are inside the threshold region, all the type-4 beads within the cutoff radius will experience an attractive force from type-3 beads. (c) when only one type-4 bead is inside threshold region, then it begins to experience bonding force with type-3 bead. (d) At this stage type-4 bead forms a strong harmonic bond with type-3 bead and this completes the phase I. The total force between type-4 bead and the type-3 bead during these stages are shown in (e), (f) and (g). Note that the force is zero in threshold region (e) as there is no qualifying beads in that region.

Equations 8 to 12 represents the heavyside functions used in the phase-I RCGMD potential, where $N_{34}$ is the number of type 4 beads around a type 3 bead, $r_{threshold}$ is the threshold distance at which we turn on the bonding potential, and $r$ is the inter-bead distance. The working principle of RCGMD phase-I potential is shown graphically in Fig. 3. Figure 3a shows the force versus inter-bead spacing for attractive, repulsive and bonding potentials. Figure 3b-d shows three cases or stages at which the bonding between type 3 and 4 happens. At stage 1 as in Fig. 3b, when all the type 4 beads are outside the threshold region of a type 3 bead, then they all experience an attractive force as shown in Fig. 3e.

When one type 4 bead enters into the threshold region, the bonding potential is turned on inside the threshold region and the repulsive potential is turned on outside it as shown in Fig. 3f. This will allow one of the type 4 bead to get bonded with type 3 bead, and other type 4 beads move away from this zone and are no longer attracted to the type 3 as seen in Fig. 3c. This process will lead to a strong bond formation between one of the type 4 and type 3 bead as shown in Fig. 3d. Again, the potential remains the same as shown in Fig. 3g. In a highly dynamic environment if multiple type 4 beads enter into the threshold region, then the bonding potential is turned off and all the type 4 beads are repelled by the type 3 bead. This rule will prevent unnecessary clustering of T-beads onto the E-region.

In a typical scenario, all type 3 beads should have a made a bond with a type 4 bead in a short period of time, to simulate the polymerization behavior of fibrin. The phase-II potentials are designed to work with fibrins which already have made type 3-4 bonds.

## Phase II: Type 2-4 bonding

In this phase of RCGMD, potentials for the polymerization of fibrin monomers is developed. The attractive, repulsive and the bonding potentials (Eqn. 14) are the same from phase-I, except the switching functions are modified this time (Eqns. 15-20). Additionally, $\Theta_6$ is introduced, which will switch on the phase-II, only if phase-I is complete. $N_{XY}$ represents the number of type $Y$ beads around type $X$ beads, and the other terms remains as the same as in phase-I.

$$V_{rcgmd-II} = \Theta_6\{(\Theta_1\Theta_4)V_{attract} + (\Theta_2\Theta_5)V_{Bonding} + (\Theta_3 + \Theta_2\Theta_4)V_{repel}\} \quad (14)$$

$$\Theta_1 = \Theta(0.9 - N_{42})\Theta(0.9 - N_{24}) \quad (15)$$

$$\Theta_2 = \left\{\Theta\left(\frac{N_{42}-0.9}{1.1-N_{42}}\right) + \Theta\left(\frac{N_{42}-1.9}{2.1-N_{42}}\right)\right\}\Theta\left(\frac{N_{24}-0.9}{1.1-N_{24}}\right) \quad (16)$$

$$\Theta_3 = \max(\Theta(N_{24}-1.9), \Theta(N_{42}-2.9)) \quad (17)$$

$$\Theta_4 = \Theta(r^2 - r_{threshold}^2) \quad (18)$$

$$\Theta_5 = \Theta(r_{threshold}^2 - r^2) \quad (19)$$

$$\Theta_6 = \Theta\left(\frac{N_{43}-0.9}{1.1-N_{43}}\right) \quad (20)$$

This potential will make sure that a bond is formed between type 2 and 4 beads, only if there is a maximum of two type 2 beads around type 4 bead, and maximum of one type 3 bead around type 4 bead. This will prevent clustering of the fibrins at a particular location.

## Computer code development

The RCGMD potentials are implemented and tested in a custom written parallel Python code. Then it was converted to a faster C++ version with which we did all the simulations in this study. The codes consist of implementation of standard potentials like Lennard Jones, harmonic bond potential and harmonic angle potential. The numerical integration is performed using velocity Verlet scheme[56], and the temperature of the system is controlled using velocity scaling thermostat. To account for large scale simulations, the code can read input from a file and data file which closely resembles that of LAMMPS[57], so that in the future the code can be easily migrated to LAMMPS[57].

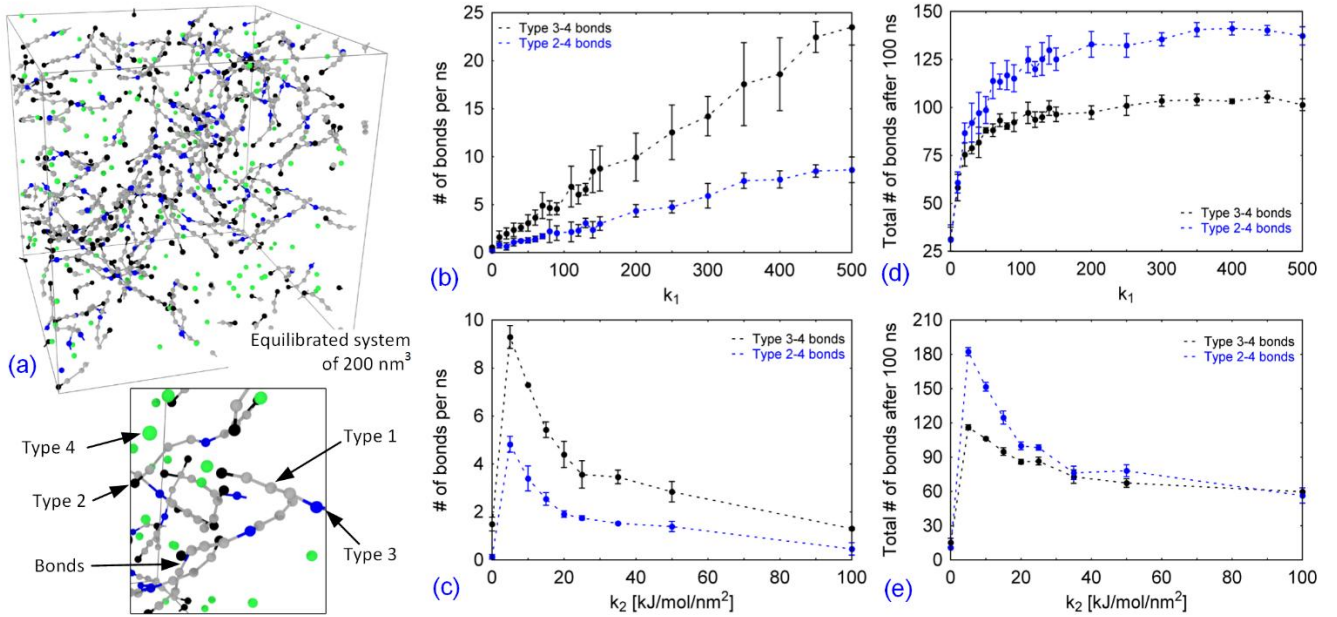

**Figure 4:** (a) A 200 $nm^3$ equilibrated molecular model with 120 fibrinogens and 150 type-4 beads is used for the simulations. The different types of beads are labeled in the subfigure. Parameter estimation and sensitivity check for $k_1$ and $k_2$ (equations 4-6) against the rate of phase-I and phase-II bonding. The figure shows number of bonds formed per ns versus (b) $k_1$ and (c) $k_2$. The total number of bonds formed after 100 ns of simulation is estimated while changing the (d) $k_1$ and (e) $k_2$.

## Results and Discussion

After the development of the RCGMD potential, the next steps were to test it to check the bonding rate, time step of integration and its influence on bonding rate and stability, parametric optimization of the potential constants etc. In this section, we will describe the simulation model that we used for the preliminary studies, optimization of the force field parameters, estimation of an optimum time step of integration and some large system fibrin polymerization studies.

### Simulation model

A cubic periodic system with sides $200\ nm$ with 120 fibrinogens and 150 type 4 beads is chosen as the molecular model for preliminary studies. This system is equilibrated for 50000 steps with a time step of $200\ fs$ before using it for any production runs. The equilibrated system is as shown in the Fig. 4a along with various types of beads marked on it. During this equilibration process, the RCGMD potentials are turned off and no bond formation is allowed with type 4 beads. The density of the fibrinogen molecules in this equilibrated system represents the physiological concentration of human fibrinogen[58] which is in the range of $3\ to\ 4.5\ g/L$. The physiological concentration of thrombin[59] is too low which makes an equivalent of three or four type 4 beads in this system. This will lead to poor polymerization, as our RCGMD is based on the central bead as type 4. Hence, we cannot directly compare thrombin with type 4, instead we should have equal number of type 4 corresponding to type 3 beads in RCGMD. To avoid the clustering of the type 4 beads into a sphere, we turned off any type of interaction among type 4 beads.

### Parameter influence on polymerization rate

To start with, the equilibration studies are performed with arbitrary values of $k_1 = 100$ and $k_2 = 15$. Our next step is to justify this selection and also the check the sensitivity of type 3-4 and 2-4 bond formation rate with respect to $k_1$ and $k_2$. Using the equilibrated model, we have performed simulation for $100\ ns$ with a time step of $200\ fs$. In the first set of simulations the $k_2 = 15$, and the $k_1$ is varied from 0 to 500. In the second set of simulations, $k_1 = 100$ and $k_2$ is varied from 0 to 100. The initial rate of bonding of type 3-4 and type 2-4 is estimated for both cases and plotted in Fig. 4b and 4c. The results show a linear correlation between the rate of bond formation and $k_1$. This trend is expected, since the increase in $k_1$ will increase the force of attraction leading to faster movement of type 4 beads towards type 3 beads, and hence bonds faster. On the other hand, second set of simulations show that increasing $k_2$ will actually decrease the chances of bonding. A range in between 0 and 20 is preferred for $k_2$, for a stable bonding as shown in Fig. 4c. This is due to the nature of the harmonic bond; as $k_2$ represents the spring constant of bonding potential, larger values of $k_2$ can lead to "bouncing off" of the type 4 bead with type 3 in conjunction with the time step of integration.

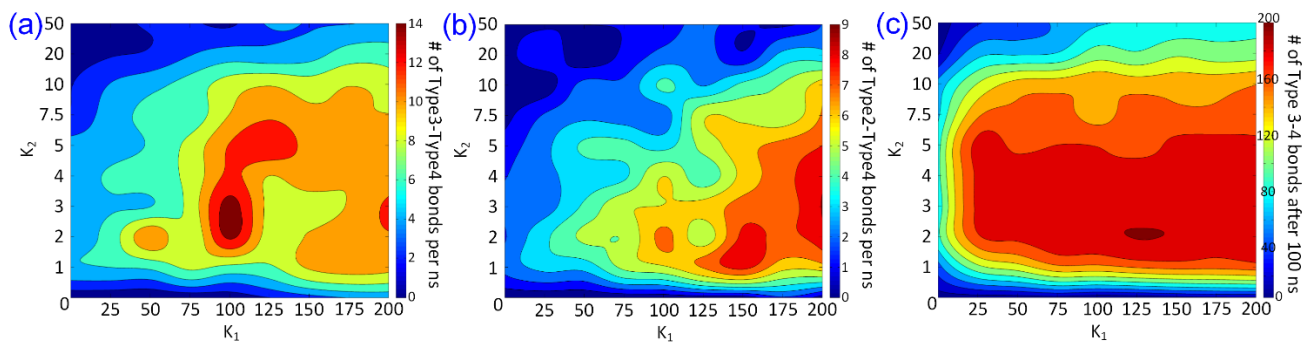

**Figure 5:** For the $200\ nm^3$ fibrinogen system, the rate of bonding is estimated by changing the parameters. (a) The variation of phase-I bonding (type-4 to type-3) with $k_1$ and $k_2$ is shown with contours representing the number of bonds per ns. (b) The variation of phase-II bonding (type-4 to type-2) with $k_1$ and $k_2$ is shown with contours representing the number of bonds per ns. (c) The total number of phase-I bonds formed over 100 ns while varying $k_1$ and $k_2$ is shown with contours representing the total number of bonds after 100 ns.

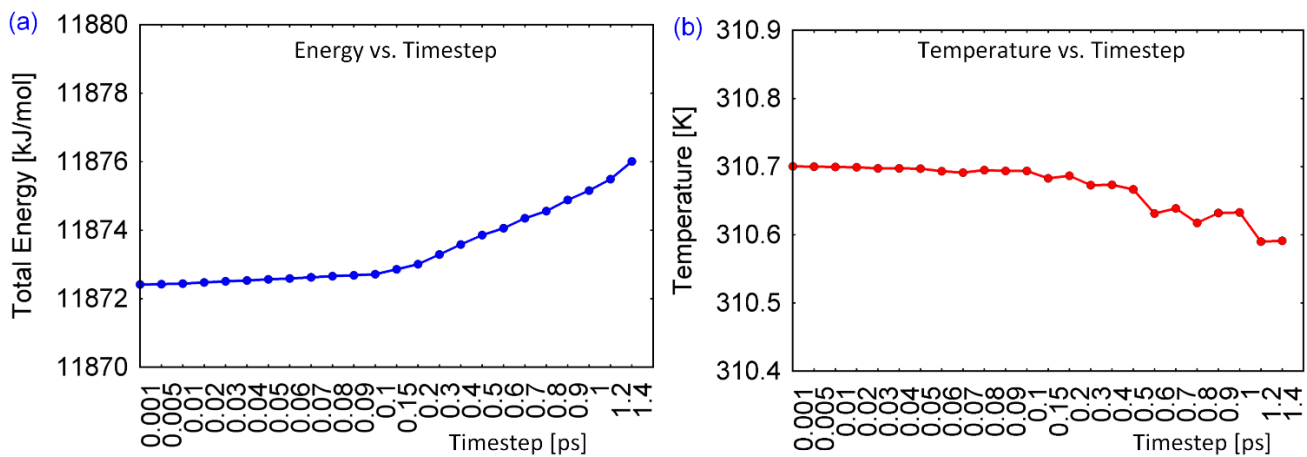

**Figure 6:** Estimation of optimal time step of integration. (a) Evolution of average total energy of the system with various time step of integration. (b) Change of system temperature with respect to time step.

The initial rate of bond formation doesn't give any information on the stability of the bonds formed. Hence, additionally, we have estimated the total number bonds formed after 100 $ns$ simulation. The results are shown in Fig. 4d and 4e for both $k_1$ and $k_2$ variations, and it shows that the $k_1$ doesn't influence much after a value of 100, but $k_2$ keeps the same trend as in Fig. 4c. So in practice, we can select $k_1$ around 100 and $k_2$ around 10. All simulations are repeated 5 times (155 in total) to reduce the statistical fluctuation and the error bars in the Fig. 4 indicate the standard deviations. After this preliminary study, we down selected the parameter ranges of $k_1$ and $k_2$ from 0 to 200 and from 0 to 50 respectively and did a two-dimensional parametric sweep to map the influence of bonding rates. The results of the 2-dimensional mapping of parameters versus bonding rate and total bond formation is shown in Fig. 5. Figure 5a and 5b shows the influence of $k_1$ and $k_2$ in type 3-4 bonding and type 2-4 bonding rates respectively. Apart from this transient bonding rate data, we estimated the influence of $k_1$ and $k_2$ on the total number of type 3-4 bonds formed after 100 $ns$ and shown in Fig. 5c. The red region shows the highest number of bond formation, with a decreasing number towards the blue region (color online). This shows that a range of 25 to 200 for $k_1$ and 1 to 15 for $k_2$ is suitable for simulating the polymerization.

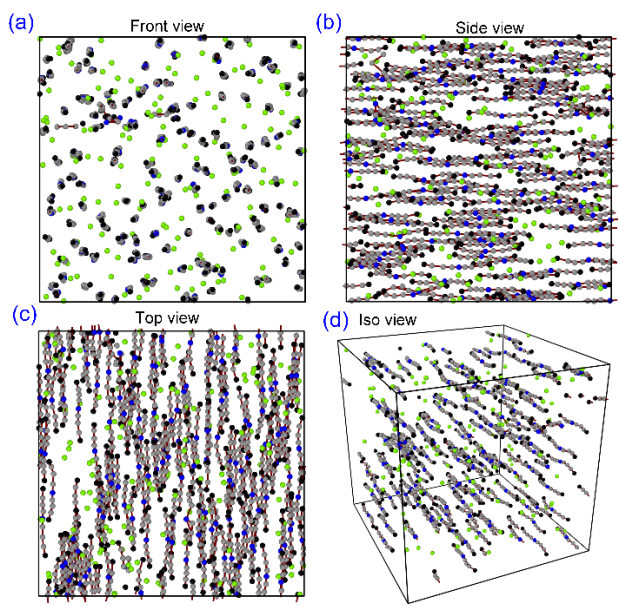

**Figure 7:** Molecular model of 150 fibrinogens and 150 type-4 beads in a 200 $nm^3$ periodic box. The figure shows the system just before equilibration. The fibrinogens and type-4 beads are inserted randomly into the box using our algorithm and the different panels show the (a) front view, (b) side view, (c) top view and (d) isometric view of the system.

**Time step of integration**

Apart from the parameters' influence, there exists other characteristics of the polymerization to explore for RCGMD simulations. One of them is the proper choice of time step of integration. Based on the results from the parametric studies, we have se-lected $k_1 = 100$ and $k_2 = 10$ and kept them constant for all RCGMD simulations going forward. For time step studies a denser system is preferred, as it can simulate fast collisions between the beads. A system with 150 fibrinogens and 250 type 4 beads in a $120\ nm^3$ cubic periodic box (not shown in figures) is chosen to study time step influence on energy and temperature. First, the system is equilibrated for $10\ ns$ at constant temperature of $310\ K$ using velocity scaling thermostat with an arbitrary time step of $200\ fs$. Then the thermostat is turned off and the time step is changed to a value ranging from $1\ fs$ to $1400\ fs$. The total energy and temperature of the system is averaged for another $10\ ps$. These values are plotted in Fig. 6a and 6b respectively. The trend shows that the average energy and temperature starts diverging when the time step value is around $200\ fs$. So an ideal choice of the time step will be $200\ fs$ or less. Going forward, our RCGMD simulations will use the time step as $200\ fs$.

**Polymerization studies**

A bigger system of $200\ nm^3$ periodic box with 150 fibrinogens and 150 type 4 beads as shown in Fig. 7 is created to study the fibrin polymerization over long periods of time. The type 4 beads and the fibrinogens are filled inside the simulation box using our random molecular filling algorithm, written in MATLAB[60]. Figure 7 shows images of this molecular model from different angles namely front view (Fig. 7a), side view (Fig. 7b), top view (Fig. 7c) and isometric view (Fig. 7d). The system is equilibrated for 50000 steps at NVT ensemble before doing any polymerization studies. The time step of integration is chosen as $100\ fs$, and the force field parameters are chosen as $k_1 = 100$ and $k_2 = 10$. The molecular dynamics simulation is performed for $2\ \mu s$ (20 million steps) to understand the molecular level polymerization details of the abovementioned fibrin system. Figure 8a-d shows the time elapsed screenshots of the fibrin polymerization using our RCGMD code.

.

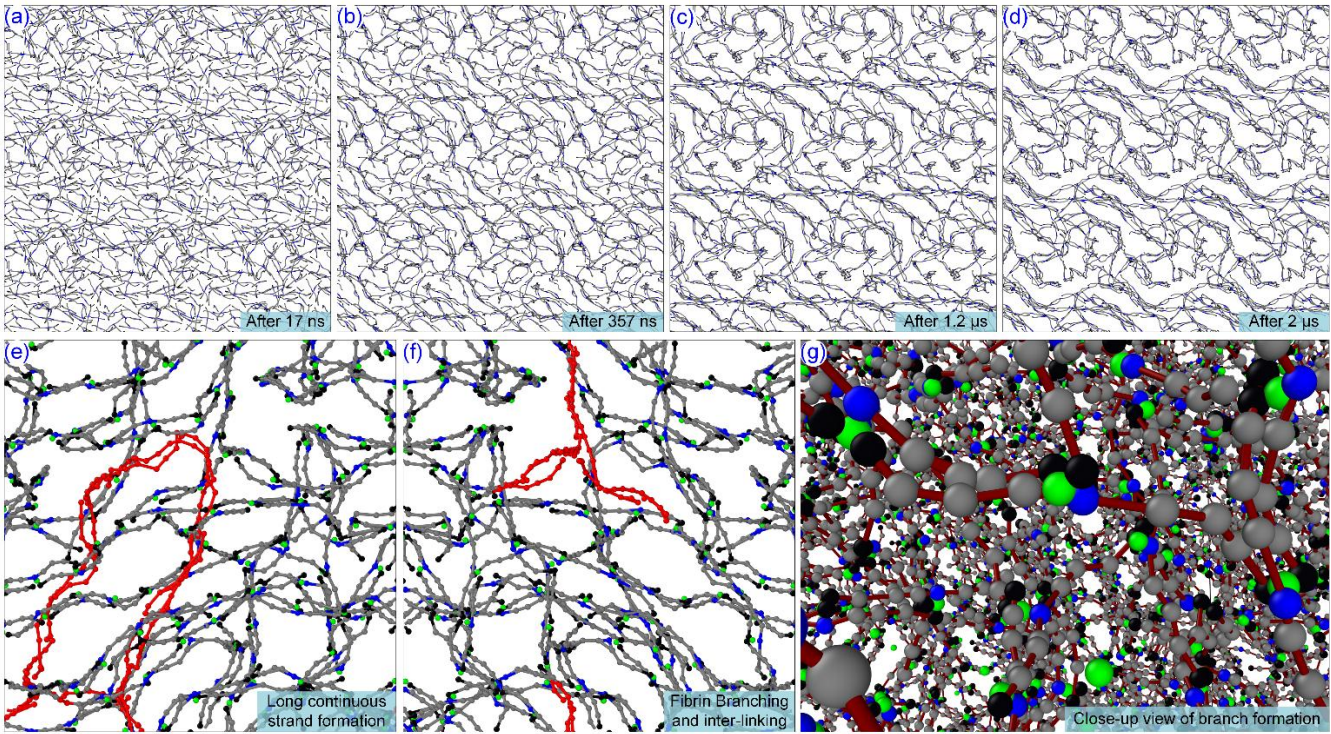

**Figure 8:** Long simulations of the fibrin system. The molecular model shown in Fig. 7 is equilibrated and run for 2μs with a time step of 100 fs. The resulting configuration is shown after (a) 17 ns, (b) 357 ns, (c) 1.2 μs and (d) 2 μs. (e) Continuous long fibrin strand formations are observed and a portion of it is marked in red. (f) Fibrin polymers started branching into two strands and is marked in red. (g) The perspective view gives a close-up view into the type4-type3-type2 interactions.

The screenshots at 17 ns (Fig. 8a), 357 ns (Fig. 8b), 1.2 µs (Fig. 8c) and 2 µs (Fig. 8d) show a gradual evolution of the fibrin polymers forming as short strands in the beginning, and then growing into long continuous networks. A portion of the long strands formed is highlighted in the Fig. 8e. After visually analyzing the domain with OVITO[61] software, we found that there is no single fibrin isolated or independent strands after 2 µs. The fibers have grown to long continuous chains and have interconnected with each other as a complex network. This is consistent with the experimental observations related to the fibrin scanning electron microscopic (SEM) studies[62]. Another important observation is the branching and cross-linking of the fibrin polymers as seen in Fig. 8f, which shows that all the branches tend to have a maximum of two branches at these intersections and is in qualitative agreement with the experimental results[62, 63]. A detailed view of the bonded atoms of the fibrinogen and type 4 beads is shown in Fig. 8g

For all the studies in this work, visualization of the molecular models is done using OVITO software[61] and molecular models are created using MATLAB[60] codes. The images shown in Fig. 8a-d is created by placing periodic images of the system along two directions, and hence they actually represent six replicas of the molecular system shown in Fig. 7.

**Scope for future improvements**

This section discusses some of the possible future improvements to be made in the RGMD model.

(i) *Quantitative validation with experimental results*: A quantitative validation can be done if we can compare the number density of the fiber junctions, thickness of the fibrin fibers, and mechanical strength of the clot. For this we need mathematical methods[64] to characterize the 3d images of fibrin clot network and then quantifying them to compare with the simulation results. This is a complex task of integrating image processing of confocal imaging, nodule (junctions of fibrin interconnections) detection algorithms, fibrin thickness assessment algorithm and fibrin density estimation algorithm.

(ii) *Mechanical properties*: Currently, we have used the spring constant as the elastic properties of the fibrinogen obtained directly from the atomic simulations[39]. However, the elastic properties of the final stable fibrin clot are not estimated and compared with the experimental studies. This needs additional simulations with stable and equilibrated fibrin clot system in conjunction with steer molecular dynamics[65] simulations.

(iii) *Energy conservation*: The switching of the RCGMD potentials around the threshold region is drastic and this may lead to poor energy conservation with the RCGMD turned on. This issue can be addressed by testing the polymerization logic with a series of different potentials along with smooth switching functions which can smoothly change the energy near the threshold radius.

(iv) *Viscosity and fluidic properties*: To extend this model to integrate with other protein models of human erythrocytes and blood plasma, we need to validate the viscosity and other fluidic properties of the fibrin system before and after clotting. This requires investigation of the system with additional particles resembling to the highly coarse-grained models of water, characterization of the potential parameters with cutoff radius etc.

(v) *Multithreaded RCGMD code*: To perform a 2 µs simulation using the current computer code written in C++ takes around 96 hours in our cluster computing as it is serial code. This can be reduced significantly by multithreading the LJ force calculation module and also utilizing faster alternate potentials for RCGMD, which can circumvent the need for square root calculation of the inter-bead distance.

We are working on the most of these challenges and will be addressed in the upcoming publication series.

## Conclusion

We have developed a reactive coarse grain molecular dynamics method to simulate the clotting mechanism of human fibrinogen. We started by developing the highly coarse-grained model of fibrinogen with only 9 beads interconnected through springs, which reflect the mechanical properties and size of the actual fibrinogen. This coarse grain model is used along with thrombin like coarse grain beads to develop our reactive coarse grain molecular dynamics (RCGMD) method. The force field parameters of this new method are characterized against the rate of polymerization and mapped into a two-dimensional space to understand the extent of their influence. Subsequently, we estimated the divergence of average total energy and temperature with different time steps of integration and found that 200 $fs$ is the best choice. Then we extended our polymerization studies with a bigger system of 150 fibrinogens and 150 type 4 beads in a 200 $nm^3$ box. These simulations predicted the time evolution of the fibrin clot formation for over 2 µs, and we have observed many real phenomena that was seen in fibrin clots. One of them is the branching of the fibrin fibers into interconnecting networks, another is the stable long continuous fibrin strand formation which are very consistent with the experimental observation. This is the first successful attempt on polymerization of the fibrin clot using molecular simulations, and we have qualitatively predicted very similar fibrin network formation observed in reality. We believe this work will be the foundation of the new class of molecular simulations called reactive coarse grain molecular dynamics, and our RCGMD concept can be optimized or tuned to accommodate any type of bio-chemical reactions at mesoscale.

## Acknowledgements

Research reported in this publication was supported by the National Heart, Lung, and Blood Institute of the National Institutes of Health under Award Number 1K01HL115486. The content is solely the responsibility of the authors and does not necessarily represent the official views of the National Institutes of Health. This study was also supported in part by resources and technical expertise from the Georgia Advanced Computing Resource Center, a partnership between the University of Georgia's Office of the Vice President for Research and Office of the Vice President for Information Technology.